\documentclass[twocolumn]{aastex631}

\shorttitle{Dynamical Stability and Habitability in the HD 20794 System}
\shortauthors{Stephen R. Kane}

\begin{document}

\title{Dynamical Stability and Habitability in the HD 20794 System}

\author[0000-0002-7084-0529]{Stephen R. Kane}
\affiliation{Department of Earth and Planetary Sciences, University of
  California, Riverside, CA 92521, USA}
\email{skane@ucr.edu}


\begin{abstract}

The Keplerian orbit of a terrestrial planet can be a significant
driver in the evolution of surface conditions, as well as influencing
the overall dynamics of the system. The HD~20794 system harbors three
confirmed planets orbiting a nearby G-type star, including HD~20794~d,
a $\sim$5.82~$M_\oplus$ (minimum mass) planet on a highly eccentric
($e = 0.45$) orbit that passes through the Habitable Zone (HZ). Here,
we present a dynamical analysis of the HD~20794 system. We calculate
the HZ boundaries and quantify the fraction of the orbital period that
planet~d spends within the conservative and optimistic HZ limits.
Using N-body simulations, we explore the long-term orbital stability
across inclinations spanning $\sim$5--90\degr. The system remains
dynamically stable over the full $10^7$~year integration for all
tested inclinations, including $i = 5\degr$ ($M_d \approx
67$~$M_\oplus$). The secular eccentricity oscillations share a common
eigenperiod that scales inversely with the total system mass,
consistent with Laplace-Lagrange secular theory. We examine the origin
of the eccentricity of planet~d, including planet-planet scattering
and secular excitation from an unseen eccentric outer
companion. HD~20794~d is the lowest-mass confirmed planet with $e >
0.4$ whose orbit crosses the HZ of its host star, and its periastron
passage deep within the HZ makes it a likely dynamical disruptor for
additional terrestrial planets, reinforcing its status as the dominant
habitability prospect in the system. The proximity of HD~20794 and its
inclusion on the Habitable Worlds Observatory precursor target list
make this a high-priority system for understanding the interplay
between orbital dynamics and planetary habitability.

\end{abstract}

\keywords{astrobiology -- planetary systems -- planets and satellites:
  dynamical evolution and stability -- stars: individual (HD~20794)}


\section{Introduction}
\label{sec:intro}

The diversity of discovered planetary system architectures has
provided extraordinary insight into orbital dynamics and the
conditions that may permit long-term planetary habitability
\citep{ford2014,winn2015,mishra2023a,mishra2023b,horner2020b,kane2021d}. A
recurring challenge in evaluating the potential habitability of
exoplanetary systems is the prevalence of eccentric orbits among
detected planets \citep{shen2008c,kane2012d,ford2008a}. Planets on
eccentric orbits that pass through the Habitable Zone (HZ) experience
time-varying stellar flux, with potentially dramatic consequences for
surface climate and atmospheric stability
\citep{williams2002,dressing2010,kane2012e,armstrong2014a,kane2017d}. The
dynamical architecture of a planetary system fundamentally constrains
the long-term viability of orbits within the HZ
\citep{kopparapu2010,kane2015b,kane2019e,kane2022a}. Giant planets on
eccentric orbits that penetrate or approach the HZ can act as
gravitational ``wrecking balls'', destabilizing the orbits of
terrestrial planets or precluding their formation entirely
\citep{kane2019e,kane2023b,kane2023c,kane2024d,kane2025c}. Even
lower-mass planets, if sufficiently eccentric, can induce secular
eccentricity oscillations in neighboring bodies that drive periodic
departures from temperate conditions
\citep{hill2018,hill2023,kane2023d,kane2024e}. Of particular interest
are those systems where a planet of ambiguous mass resides on an
eccentric orbit within the HZ, as such configurations may range from
being promising candidates for habitability to dynamical disruptors,
depending on the unknown orbital inclination and true planetary mass.

The HD~20794 system is an interesting case study for investigating
these questions. At a distance of only 6.04~pc, HD~20794 is among the
nearest G-type main-sequence stars known to host confirmed planets
\citep{pepe2011,nari2025}. The system contains three confirmed radial
velocity (RV) detected planets: HD~20794~b ($P = 18.3$~d, $M_p \sin i
= 2.15$~$M_\oplus$), HD~20794~c ($P = 89.7$~d, $M_p \sin i =
2.98$~$M_\oplus$), and HD~20794~d ($P = 647.6$~d, $M_p \sin i =
5.82$~$M_\oplus$) \citep{nari2025}. The inner two planets are
consistent with near-circular orbits, but planet~d exhibits a
substantial eccentricity of $e = 0.45^{+0.10}_{-0.11}$, placing it on
an orbit that traverses the stellar HZ. The planetary nature of the
647.6~d signal was initially detected by \citet{cretignier2023} using
HARPS data processed with the YARARA pipeline, and subsequently
confirmed by \citet{nari2025} using combined HARPS and ESPRESSO
observations spanning more than 20 years. Despite the considerable
scientific interest in this system, no comprehensive dynamical study
has been conducted to date, leaving the orbital stability and
habitability implications of the system unexplored.

In this paper, we present a full dynamical analysis of the HD~20794
planetary system, with a focus on the eccentric orbit of planet~d and
its implications for habitability. Section~\ref{sec:arch} describes
the stellar and planetary properties and the calculation of the HZ
boundaries. The description of the N-body simulation methodology and
results of the stability analysis are presented in
Section~\ref{sec:dyn}. In Section~\ref{sec:ori} we investigate the
origin of the eccentricity of planet~d. The implications for
habitability and the relevance to direct imaging missions are
discussed in Section~\ref{sec:disc}, and we provide concluding remarks
in Section~\ref{sec:con}.


\section{System Architecture and Habitable Zone}
\label{sec:arch}

Here we describe the HD~20794 system architecture and calculate the
HZ boundaries.


\subsection{Host Star}
\label{sec:host}

HD~20794 (82~G.~Eridani, GJ~139, HIP~15510) is a bright ($V = 4.34$),
nearby ($d = 6.04 \pm 0.01$~pc) G-type main-sequence star in the
constellation Eridanus. We adopt the stellar parameters from the
analysis of \citet{nari2025}, summarized in Table~\ref{tab:stellar}.
The star has an effective temperature of $T_\mathrm{eff} = 5368 \pm
48$~K, a luminosity of $L_\star = 0.687 \pm 0.003$~$L_\odot$, a mass
of $M_\star = 0.79 \pm 0.01$~$M_\odot$, and a radius of $R_\star =
0.93 \pm 0.03$~$R_\odot$. A notable characteristic of HD~20794 is its
subsolar metallicity ($[\mathrm{Fe/H}] \approx -0.42$), which places
it among the more metal-poor planet-hosting stars
\citep{buchhave2014,brewer2018b}. Age estimates are uncertain, ranging
from $5.76 \pm 0.66$~Gyr based on chromospheric activity indicators to
$14^{+5}_{-6}$~Gyr from evolutionary track fitting
\citep{nari2025}. The low level of stellar activity, as evidenced by a
magnetic cycle of $\sim$3000~d and a rotation period of $\sim$39~d
\citep{nari2025},
confirms the suitability of HD~20794 for high-precision RV studies. A
cold debris disk has been detected at $\sim$24~AU via Herschel
observations \citep{kennedy2015d}, suggestive of a reservoir of icy
material in the outer system. HD~20794 is included on the Habitable
Worlds Observatory (HWO) Exoplanet Exploration Program precursor
target list \citep{mamajek2024,tuchow2024}, underscoring its
importance for future direct imaging characterization efforts
\citep{kopparapu2018,laliotis2023,harada2024b}.

\begin{deluxetable}{lc}
\tablecaption{\label{tab:stellar} Stellar parameters of HD~20794.}
\tablehead{
  \colhead{Parameter} &
  \colhead{Value}
}
\startdata
  Spectral Type       & G6V                    \\
  $V$ (mag)           & 4.34                   \\
  Distance (pc)       & $6.04 \pm 0.01$        \\
  $T_\mathrm{eff}$ (K) & $5368 \pm 48$        \\
  $L_\star$ ($L_\odot$) & $0.687 \pm 0.003$   \\
  $M_\star$ ($M_\odot$) & $0.79 \pm 0.01$     \\
  $R_\star$ ($R_\odot$) & $0.93 \pm 0.03$     \\
  $[\mathrm{Fe/H}]$   & $-0.42 \pm 0.02$     \\
  Age (Gyr)           & $5.76 \pm 0.66$        \\
  $P_\mathrm{rot}$ (d) & $\sim$39             \\
\enddata
\tablecomments{All values adopted from \citet{nari2025}.}
\end{deluxetable}


\subsection{Planetary Parameters}
\label{sec:planets}

\begin{deluxetable*}{lccc}
\tablecolumns{4}
\tablecaption{\label{tab:planets} Planetary parameters of the
  HD~20794 system.}
\tablehead{
  \colhead{Parameter} &
  \colhead{HD~20794~b} &
  \colhead{HD~20794~c} &
  \colhead{HD~20794~d}
}
\startdata
  $P$ (d)              & $18.3142 \pm 0.0022$   & $89.68 \pm 0.10$     & $647.6^{+2.5}_{-2.7}$    \\
  $K$ (m\,s$^{-1}$)   & $0.614 \pm 0.048$      & $0.502^{+0.048}_{-0.049}$    & $0.567^{+0.067}_{-0.064}$         \\
  $M_p \sin i$ ($M_\oplus$) & $2.15 \pm 0.17$    & $2.98 \pm 0.29$      & $5.82 \pm 0.57$           \\
  $a$ (AU)             & $0.121 \pm 0.001$ & $0.350 \pm 0.002$ & $1.354 \pm 0.007$   \\
  $e$                  & $0.064^{+0.065}_{-0.046}$ & $0.077^{+0.084}_{-0.055}$ & $0.45^{+0.10}_{-0.11}$    \\
\enddata
\tablecomments{Values adopted from \citet{nari2025}. Semi-major axis uncertainties are estimated through error propagation.}
\end{deluxetable*}

The planetary system of HD~20794 has a complex detection history that
has evolved substantially over the past 15 years as RV precision and
data analysis techniques have improved. \citet{pepe2011} reported the
initial discovery of three super-Earth candidates from RV
observations, with orbital periods of 18.3, 40.1, and 90.3~d and
minimum masses of 2.7, 2.4, and 4.8~$M_\oplus$, respectively,
designated planets~b, c, and~d. A subsequent re-analysis by
\citet{feng2017a}, employing a differential RV technique, found
evidence for candidates at periods of 18, 89, 147, and 330~d, with
weak evidence for additional signals at 43~d and 11.9~d, but did not
confirm the 40.1~d planet reported by \citet{pepe2011}. Similarly, the
archival RV survey of \citet{laliotis2023} confirmed only the 18~d and
90~d planets. The non-detection of the 40.1~d signal in these later
analyses, despite the increased data baseline, cast significant doubt
on its planetary origin. The period of this signal is notably close to
the stellar rotation period of $\sim$39~d, suggesting that it may
instead arise from surface activity modulation
\citep{desort2007,robertson2014a,nari2025}.

A major advance was produced by \citet{cretignier2023}, who confirmed
the 18~d and 89~d planets and detected a new candidate signal at a
period of $\sim$650~d. This long-period candidate was subsequently
confirmed by \citet{nari2025}, who conducted a comprehensive analysis
combining more than 20 years of RV observations. \citet{nari2025}
compared models with zero through four planets and found that the
three-planet model (at periods of 18.3, 89.7, and 647.6~d) was
strongly preferred. The additional planet candidates reported by
\citet{feng2017a} at periods of 147 and 330~d were not recovered. We
adopt the planetary parameters and nomenclature from \citet{nari2025},
summarized in Table~\ref{tab:planets}, wherein the three confirmed
planets are designated b, c, and d in order of increasing orbital
period. HD~20794~d is the outermost confirmed planet and the primary
focus of this investigation. The planet has a minimum mass of $M_d
\sin i = 5.82$~$M_\oplus$, a semi-major axis of $a_d \approx
1.354$~AU, and an eccentricity of $e_d = 0.45$, yielding a periastron
and apastron distances of 0.74~AU and 1.96~AU, respectively. Note that
\citet{nari2025} do not provide the periastron arguments for the
planetary orbits, so we assume values of 90\degr for the subsequent
analyses.


\subsection{Habitable Zone}
\label{sec:hz}

\begin{figure*}
  \begin{center}
    \begin{tabular}{cc}
      \includegraphics[angle=270,width=\columnwidth]{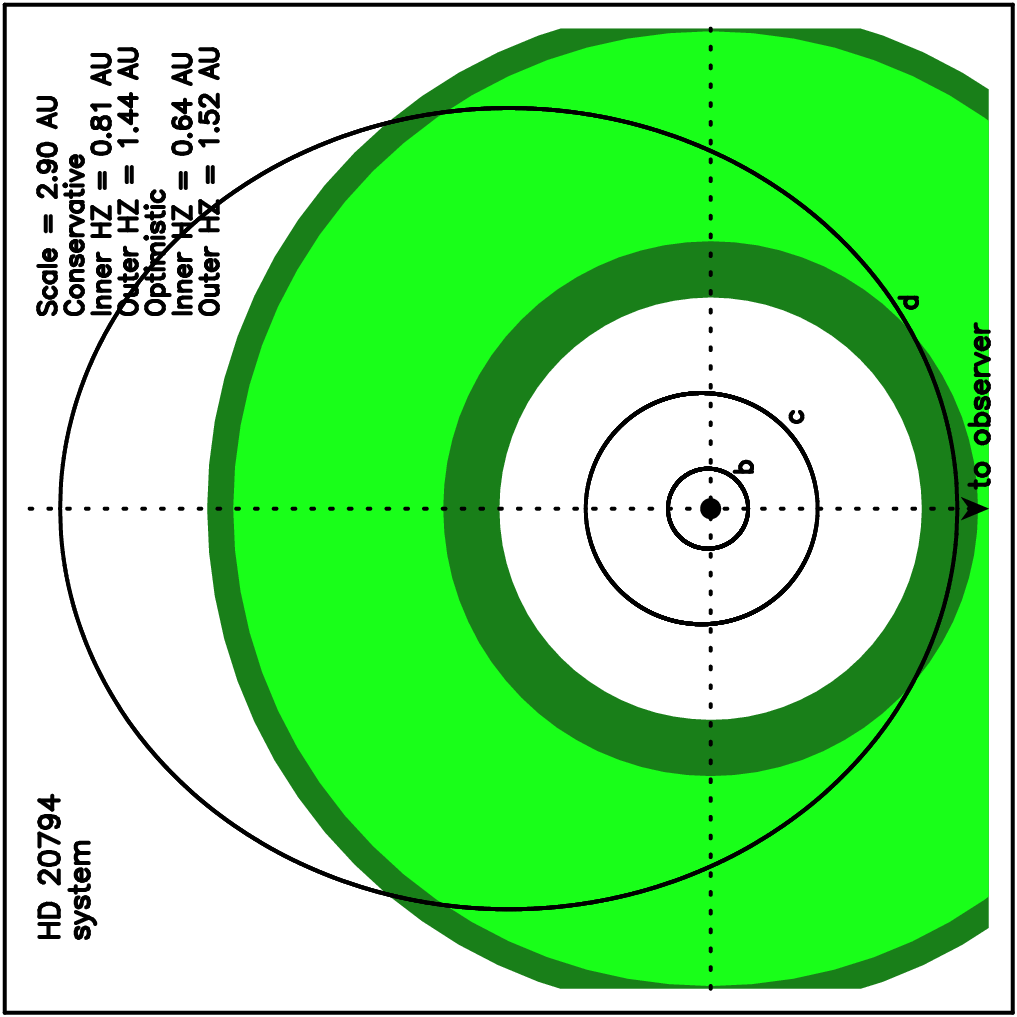} &
      \includegraphics[angle=270,width=\columnwidth]{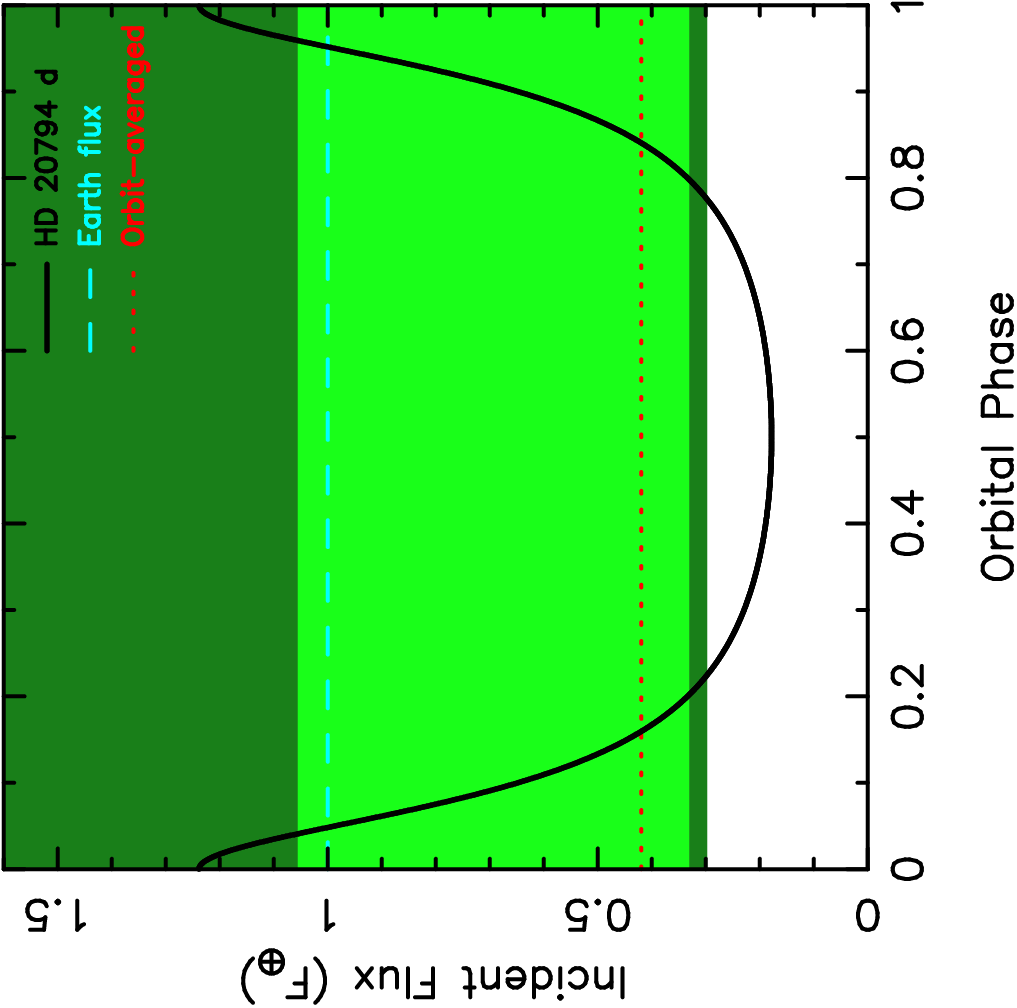}
    \end{tabular}
  \end{center}
  \caption{Left: HZ and planetary orbits in the HD~20794 system, where
    the orbits are labeled by planet designation. The scale of the
    figure is 2.9~AU along each side. Right: Variation in incident
    flux received by HD~20794~d during one complete orbit. For both
    panels, the extent of the HZ is shown in green, where light green
    and dark green indicate the CHZ and OHZ, respectively.}
  \label{fig:hz}
\end{figure*}

We calculate the HZ boundaries of HD~20794 using the formalism of
\citet{kopparapu2013a,kopparapu2014}. The conservative HZ (CHZ),
defined by the runaway greenhouse and maximum greenhouse limits, spans
from 0.807~AU to 1.440~AU ($S_\mathrm{eff} = 1.055$--0.331~$F_\oplus$).
The optimistic HZ (OHZ), defined by the
recent Venus and early Mars empirical limits, extends from 0.637~AU to
1.519~AU ($S_\mathrm{eff} = 1.693$--0.298~$F_\oplus$). The left panel of Figure~\ref{fig:hz} shows a top-down view
of the system, including the planetary orbits overlaid on the CHZ
(dark green) and OHZ (light green). The eccentric orbit of HD~20794~d
carries the planet across a wide range of stellar distances during
each orbital period, resulting in the planet transiting both the inner
and outer boundaries of the HZ, spending only a fraction of each orbit
within the temperate zone. Following the methodology described in
\citet{kane2012e} and \citet{kane2021a}, we compute the fraction of
the orbital period spent within the HZ boundaries, weighted by the
time the planet spends at each orbital position. For the CHZ,
HD~20794~d spends approximately 32\% of its orbital period within the
temperate zone, increasing to approximately 45\% for the OHZ. The
remaining $\sim$55\% of the orbit is spent exterior to the OHZ outer
boundary, where the reduced stellar flux may drive surface
temperatures below the freezing point of water for plausible
atmospheric compositions \citep{kane2012e}.

The orbit-averaged incident flux provides an alternative metric for
assessing the habitability of planets on eccentric orbits,
particularly in terms of climate forcing
\citep{williams2002,kane2021a}. The time-averaged flux is given by:
\begin{equation}
  \langle F \rangle = \frac{L_\star}{4\pi a^2 (1 - e^2)^{1/2}} \; ,
\end{equation}
which yields $\langle F \rangle \approx 0.42$~$F_\oplus$ for
HD~20794~d, where $F_\oplus$ is the flux received by Earth. Propagating the uncertainties on $a$ and $e$ from
Table~\ref{tab:planets} yields $\langle F \rangle =
0.42^{+0.03}_{-0.02}$~$F_\oplus$. The
variation in the incident flux, relative to both the time-averaged
and Earth fluxes, are depicted in the right panel of
Figure~\ref{fig:hz}. The time-averaged flux falls within the CHZ,
suggesting that on average the planet receives an insolation level
consistent with temperate conditions. However, the instantaneous flux
varies from $\sim$1.24~$F_\oplus$ at periastron to
$\sim$0.18~$F_\oplus$ at apastron, a factor of $\sim$7 variation that
may drive extreme seasonal cycles. The periastron flux is
particularly sensitive to the eccentricity uncertainty, with a
1$\sigma$ range of $\sim$0.9--1.9~$F_\oplus$, while the apastron flux
is better constrained at $0.18 \pm 0.03$~$F_\oplus$.


\section{Dynamical Analysis}
\label{sec:dyn}

Here we describe the N-body simulation methodology and present the
results of the dynamical analysis.


\subsection{Simulation Methodology}
\label{sec:methods}

\begin{figure}
  \includegraphics[angle=270,width=\linewidth]{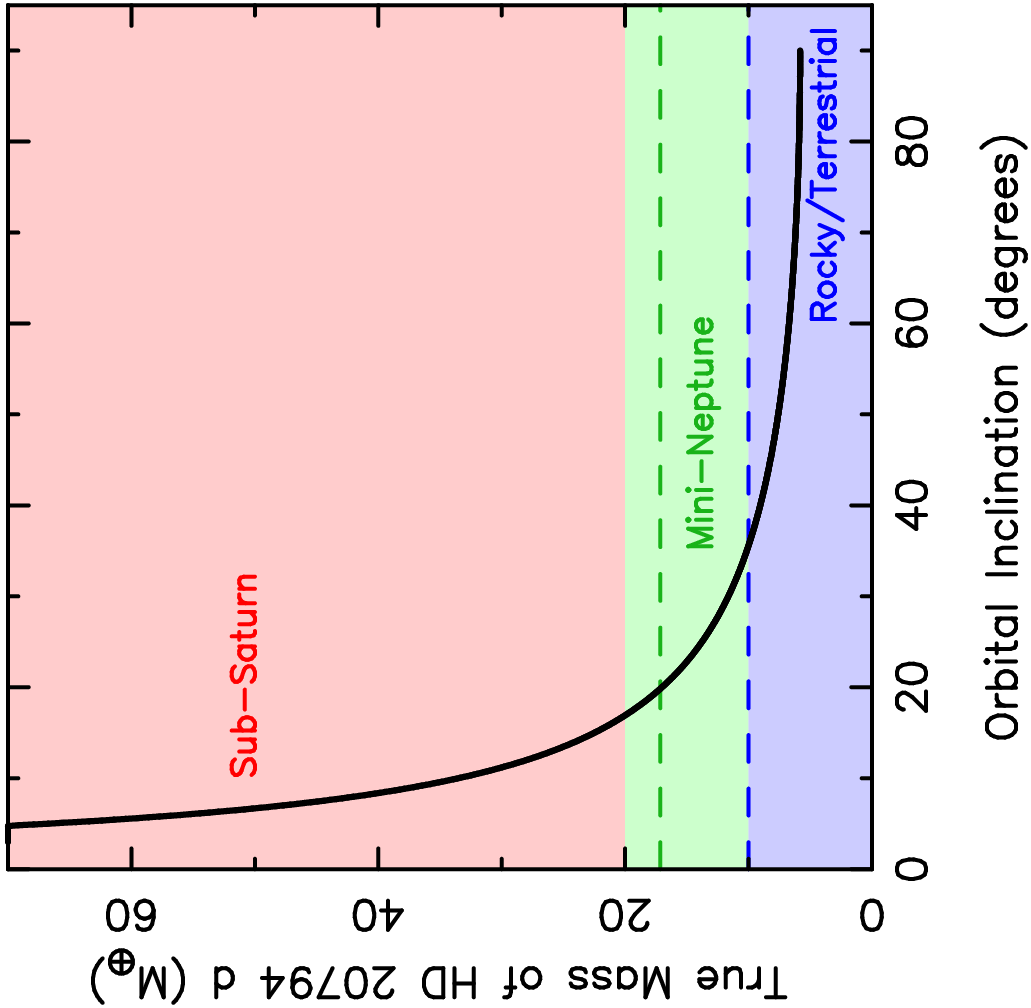}
  \caption{The true mass of planet~d as a function of assumed orbital
    inclination. The shaded regions indicate the likely planetary
    nature, including terrestrial, mini-Neptune, and sub-Saturn.}
  \label{fig:massinc}
\end{figure}

We conducted N-body simulations of the HD~20794 system using the
Mercury Integrator Package \citep{chambers1999} with a hybrid
symplectic/Bulirsch-Stoer integrator and a Jacobi coordinate system
\citep{wisdom1991,wisdom2006b}, following the methodology described by
\citet{kane2019e,kane2021a,kane2023b}. The simulations were
initialized using the orbital elements from \citet{nari2025}
(Table~\ref{tab:planets}), including the best-fit eccentricities. All simulations assume coplanar, prograde orbits, with the three
planets sharing a common orbital inclination $i$ with respect to the
line of sight. All
of the simulations were integrated for $10^7$~years with a timestep of
0.1~days, orbital elements were output every 100~years, and planetary
collisions and ejections were recorded. A central goal of this study
is to explore the dependence of the system stability on the true
masses of the planets, which are unconstrained due to the unknown
orbital inclination. We parameterize this by running suites of
simulations across a grid of assumed inclinations: $i = 90^\circ$,
30$^\circ$, 20$^\circ$, 10$^\circ$, and 5$^\circ$. At each
inclination, the true masses of all three planets are scaled as $M = M
\sin i / \sin(i)$. The corresponding true masses of planet~d range
from $M_d \approx 5.82$~$M_\oplus$ (at $i = 90^\circ$) to $M_d \approx
66.8$~$M_\oplus$ (at $i = 5^\circ$), with the transition from the
super-Earth to the Neptune-mass regime occurring at inclinations of
$\sim$20$^\circ$ ($M_d \sim 17$~$M_\oplus$). Empirical studies of the
mass-radius relation for exoplanets have shown that the transition
from predominantly rocky compositions to planets with substantial
volatile envelopes occurs at $\sim$4--10~$M_\oplus$
\citep{weiss2014,rogers2015a,chen2017,otegi2020,luque2022b,muller2024b},
above which planets are increasingly likely to retain thick H/He or
water-rich atmospheres. The mass of Neptune (17.15~$M_\oplus$)
provides a natural upper reference point, above which a planet would
unambiguously reside in the ice giant regime. The true mass of
planet~d as a function of assumed inclination, along with regions of
likely planetary nature, are represented in Figure~\ref{fig:massinc}.

\subsection{Stability as a Function of Inclination}
\label{sec:stability}

\begin{figure*}
  \includegraphics[angle=270,width=\linewidth]{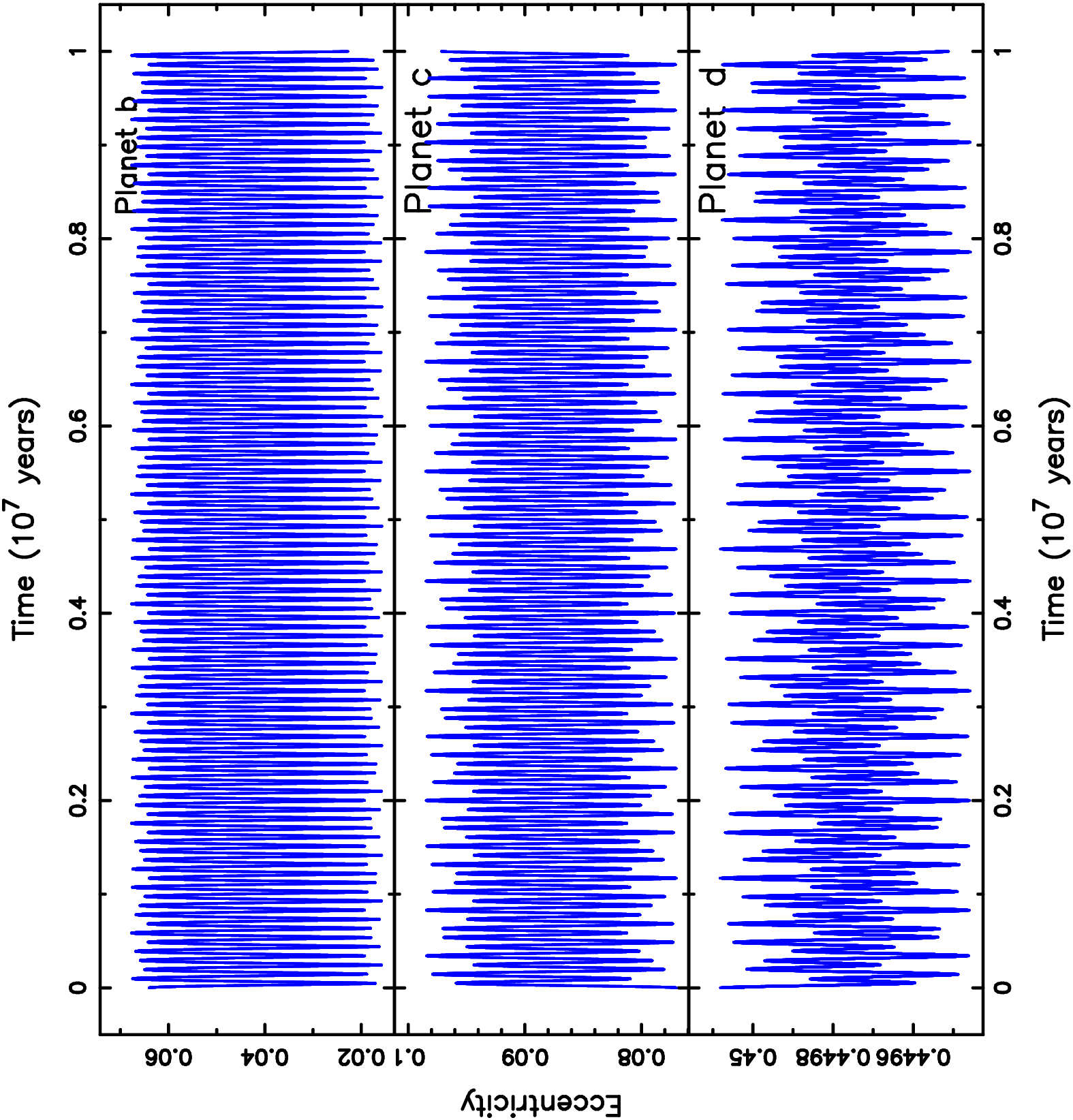}
  \caption{Eccentricity evolution for the three known planets of the
    HD~20794 system for the inclination case of $i = 90\degr$. Data
    are shown for the full $10^7$ year integration.}
  \label{fig:ecc}
\end{figure*}

The results of the inclination-dependent stability analysis
demonstrate that the HD~20794 system is remarkably robust to changes
in the assumed planetary masses. For all tested inclinations, from $i
= 90\degr$ ($M_d \approx 5.82$~$M_\oplus$) down to $i = 5\degr$ ($M_d
\approx 66.8$~$M_\oplus$), the three-planet system remains dynamically
stable over the full $10^7$~year integration, with no planetary
ejections or collisions. For example, shown in Figure~\ref{fig:ecc}
are the simulation results for the $i = 90\degr$ case, demonstrating
the eccentricity variations and relative stability over $10^7$~years
for each of the three planets in the system. Additional tests at
inclinations as low as $i = 1\degr$ (corresponding to true masses
exceeding 300~$M_\oplus$ for planet~d) also yielded stable
configurations. This stability reflects the large dynamical separation
between the planets: the ratio of the semi-major axes of planets~c
and~d is $a_c / a_d \approx 0.26$, and the ratio for planets~b and~c
is $a_b / a_c \approx 0.35$, both of which place the planets well
beyond the critical spacing for orbital instability
\citep{gladman1993,chambers1996,barnes2004b}. The wide spacing, even
combined with the substantial eccentricity of planet~d, prevents orbit
crossing for all tested mass configurations. This result demonstrates
that the true mass of planet~d is not constrained by the dynamical
stability of the system, in contrast to many other multi-planet
systems where instability provides an upper limit on the orbital
inclination \citep{kane2019e,kane2023c,kane2025c}. We note that
these stability conclusions apply to the coplanar case; the mutual
inclinations of the planets are unconstrained by RV data. However,
given the wide orbital separations, modest mutual inclinations
($\lesssim$10--20$^\circ$) are unlikely to alter the stability
picture significantly.

\subsection{Eccentricity Evolution and Secular Dynamics}
\label{sec:secular}

Although the system remains stable at all inclinations, the secular
evolution of the eccentricities reveals a rich dynamical structure
that is strongly modulated by the total system mass. In all
simulations, the eccentricities of the three planets undergo coupled
periodic oscillations driven by the secular gravitational interaction
among the planets \citep{murray1999a,laskar2004c}. For the
minimum-mass configuration ($i = 90^\circ$), the eccentricity of
planet~b oscillates between $\sim$0.016 and $\sim$0.068 with a mean of
$\sim$0.045, while planet~c oscillates between $\sim$0.077 and
$\sim$0.099 with a mean of $\sim$0.088. The eccentricity of planet~d,
by contrast, oscillates over a very narrow range near its initial
value of 0.45, with a total amplitude of only $\sim$0.0006 (see
Figure~\ref{fig:ecc}). This behavior is characteristic of
Laplace-Lagrange secular theory, where the much larger mass and
eccentricity of planet~d cause it to dominate the secular eigenmodes
while experiencing minimal back reaction from the lower-mass inner
planets \citep{murray1999a,petrovich2013}.

\begin{figure}
  \includegraphics[angle=270,width=\columnwidth]{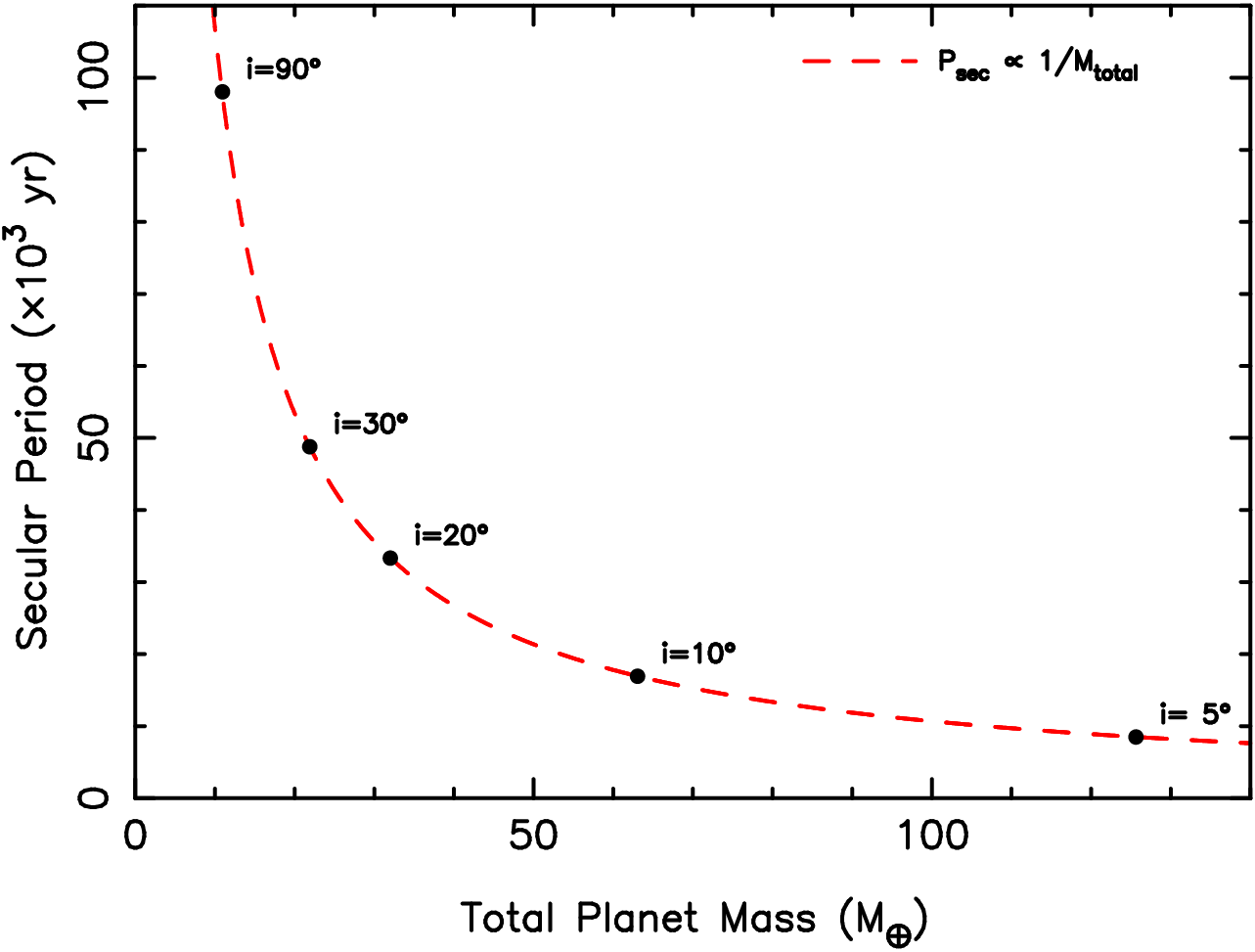}
  \caption{The dominant secular eigenperiod of the eccentricity
    oscillations as a function of total planet mass for the HD~20794
    system. The red dashed line shows the $P_\mathrm{sec} \propto
    1/M_\mathrm{total}$ scaling predicted by Laplace-Lagrange secular
    theory, which matches the simulation results to within 0.7\%.}
  \label{fig:secperiod}
\end{figure}

An interesting feature of the simulations is that all three planets
share a common dominant secular eigenperiod at each inclination, and
this eigenperiod decreases systematically with increasing total system
mass. For example, at $i = 90\degr$ (total mass
$\sim$11.0~$M_\oplus$), the dominant eccentricity oscillation period
is $\sim$98,000~years. At $i = 30\degr$ ($\sim$21.9~$M_\oplus$), the
period decreases to $\sim$49,000~years, and at $i = 10\degr$
($\sim$63.1~$M_\oplus$), the period decreases to
$\sim$17,000~years. The dominant eccentricity oscillation period for
all tested inclinations down to $i = 5\degr$ are shown in
Figure~\ref{fig:secperiod}. The product of the secular period
$P_\mathrm{sec}$ and total system mass $M_\mathrm{total}$ is constant
to within 0.7\% across all simulations, confirming the linear scaling
$P_\mathrm{sec} \propto 1/M_\mathrm{total}$ predicted by
Laplace-Lagrange secular theory \citep{murray1999a,laskar2008}. In the
Laplace-Lagrange framework, the secular eigenfrequencies are
proportional to the planetary masses divided by the stellar mass and
the semi-major axis ratios, so that scaling all planet masses by a
common factor (as occurs when changing the system inclination)
produces a proportional increase in the eigenfrequencies
\citep{murray1999a}. The precise adherence of our simulations to this
scaling confirms that the system dynamics are well described by linear
secular theory over the mass range explored. Furthermore, the
  semi-major axes of all three planets remain effectively constant
  throughout the $10^7$~year integrations at all tested inclinations,
  with variations of less than $\sim$0.01\%. This is consistent with
  the conservation of semi-major axes in Laplace-Lagrange secular
  theory, where only the eccentricities and longitudes of periastron
  undergo periodic evolution \citep{murray1999a}. The negligible
  semi-major axis variability implies that the instellation variations
  discussed in Section~\ref{sec:hz} are driven entirely by the
  eccentricity oscillations of planet~d, and do not change
  significantly as a function of system age.

\subsection{Forced Eccentricity and Inclination Constraints}
\label{sec:forced}

Although the dynamical stability of the system does not constrain the
orbital inclination, the amplitude of the forced eccentricity
oscillations of the inner planets provides an indirect observational
diagnostic. The eccentricity amplitude of planet~b decreases modestly
from $\Delta e_b \approx 0.052$ at $i = 90^\circ$ to $\Delta e_b
\approx 0.049$ at $i = 5^\circ$, and similarly for planet~c ($\Delta
e_c \approx 0.021$ to 0.019). These amplitudes remain consistent with
the best-fit eccentricities of planets~b ($e_b =
0.064^{+0.065}_{-0.046}$) and~c ($e_c = 0.077^{+0.084}_{-0.055}$)
reported by \citet{nari2025} at all tested inclinations. The
near-constancy of these amplitudes across the mass range is a
consequence of the secular eigenmode structure: the forced
eccentricity of the inner planets is set primarily by the eccentricity
of planet~d and the semi-major axis ratios, with only a weak
dependence on the absolute masses when all masses are scaled together
\citep{murray1999a}. The eccentricity amplitude of planet~d itself
increases modestly from $\Delta e_d \approx 0.0006$ at $i = 90^\circ$
to $\Delta e_d \approx 0.0010$ at $i = 5^\circ$, reflecting the
increased back reaction from the more massive inner planets.

The measured eccentricities of planets~b and~c are
sufficiently uncertain that they do not currently constrain the system
inclination. However, future observations with improved RV precision
that tighten the eccentricity constraints on the inner planets could
in principle probe the secular eigenmode structure and provide
independent constraints on the system architecture. If the
eccentricities of planets~b and~c are measured to be significantly
below the oscillation amplitudes predicted by the simulations, this
would suggest additional dissipative processes (e.g., tidal damping)
or a different orbital configuration than assumed here.


\section{Origin of the Eccentricity}
\label{sec:ori}

The relatively high eccentricity of HD~20794~d is notable for a planet
with a minimum mass in the super-Earth regime, as such planets are
typically found on near-circular orbits
\citep{ford2008a,shen2008c,kane2012d,vaneylen2015}. Several mechanisms
could produce or maintain this eccentricity, which we investigate
here.


\subsection{Planet-Planet Scattering}
\label{sec:scattering}

Planet-planet scattering is one of the most widely invoked mechanisms
for producing eccentric orbits in exoplanetary systems
\citep{rasio1996c,chatterjee2008,juric2008b,ford2008c}. In this
scenario, gravitational interactions between two or more planets lead
to close encounters that result in the ejection of one or more bodies
and the excitation of the surviving planets onto eccentric orbits. If
HD~20794 originally hosted an additional planet between the present
orbits of planets~c and~d, a scattering event could have removed the
interloper while pumping the eccentricity of planet~d. The relatively
low minimum mass of planet~d makes it an atypical candidate for the
scattering scenario, which more commonly involves giant planets
\citep{chatterjee2008,ford2008c,juric2008b,raymond2009b,carrera2019b}. However,
scattering among super-Earth-mass bodies has been demonstrated in
numerical simulations \citep{raymond2009c}, and the low metallicity of
HD~20794 may have favored the formation of multiple super-Earths
rather than a single giant planet.

\begin{figure}
  \includegraphics[angle=270,width=\linewidth]{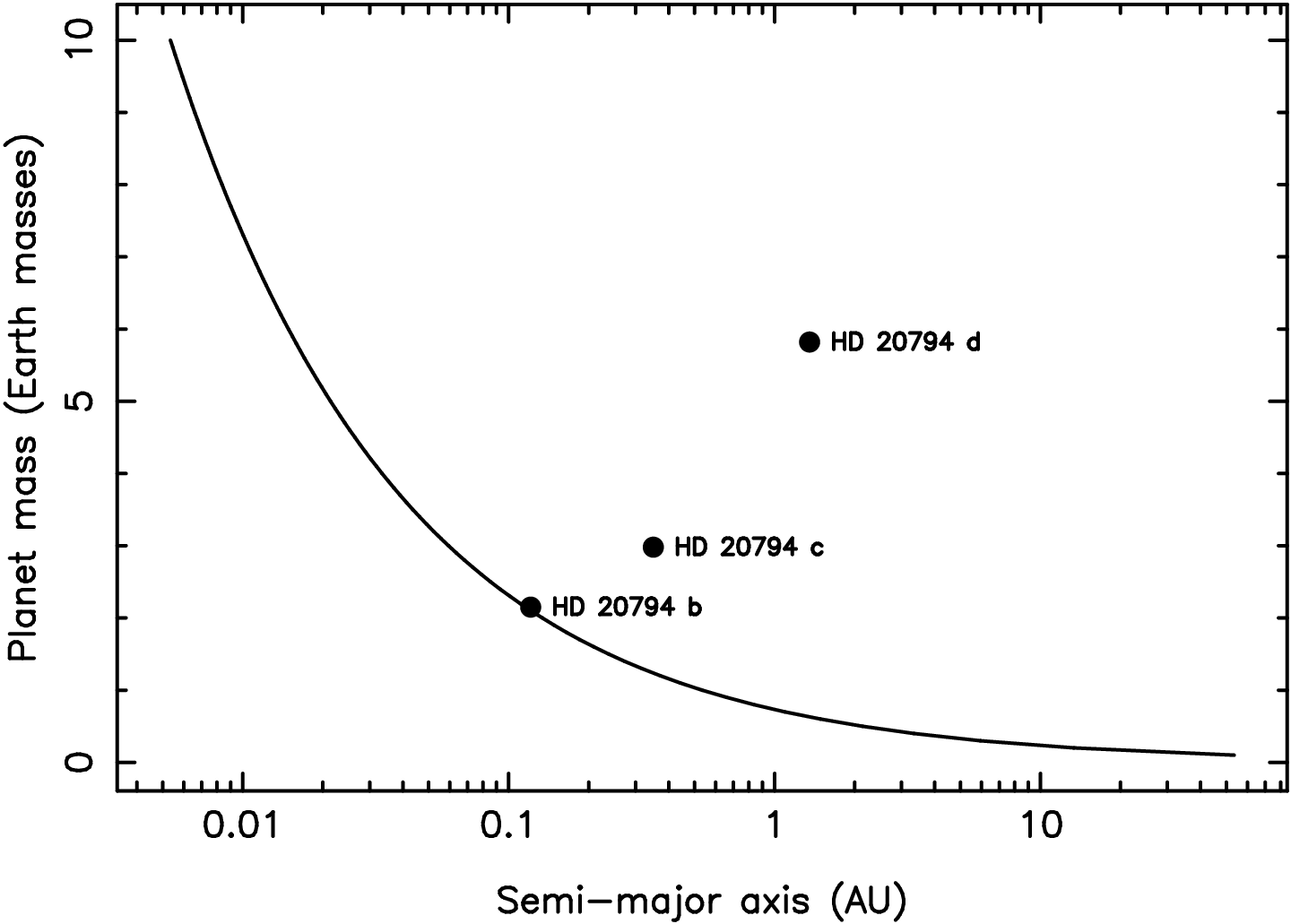}
  \caption{The mass and semi-major axis of an additional planet in a
    circular orbit whose angular momentum equals the angular momentum
    deficit (AMD) for the HD~20794 system, and may have been
    previously ejected from the system. The dots indicate the minimum
    masses and semi-major axes of the known planets in the system.}
  \label{fig:amd}
\end{figure}

A useful diagnostic for assessing the degree of dynamical excitation
is the angular momentum deficit (AMD), which quantifies the difference
between the angular momentum of the actual system and that of the
equivalent circular, coplanar configuration
\citep{laskar1997,laskar2017,murray1999a}. For the minimum-mass,
coplanar configuration of HD~20794 based on the parameters in
Table~\ref{tab:planets}, the circular angular momenta of the three
planets are $\Lambda_b = 1.77 \times 10^{40}$, $\Lambda_c = 4.17
\times 10^{40}$, and $\Lambda_d = 1.60 \times 10^{41}$
kg~m$^2$~s$^{-1}$, yielding a total circular angular momentum of
$\Lambda_\mathrm{tot} = 2.20 \times 10^{41}$~kg~m$^2$~s$^{-1}$. For
comparison, the angular momentum values for Venus and Earth are $1.84
\times 10^{40}$ and $2.66 \times 10^{40}$~kg~m$^2$~s$^{-1}$,
respectively. The corresponding HD~20794 system AMD is $1.73 \times
10^{40}$~kg~m$^2$~s$^{-1}$, or $\approx 7.9\%$ of the total circular
angular momentum budget. The AMD is overwhelmingly dominated by
planet~d, which contributes $\approx 99.1\%$ of the total AMD. Because
these calculations adopt the RV minimum masses, both $\Lambda_k$ and
the AMD scale approximately as $1/\sin i$ for a common coplanar
inclination, while the fractional dominance of planet~d remains
essentially unchanged. Figure~\ref{fig:amd} shows the mass and
semi-major axis of an additional planet (on a circular orbit) that
alone would produce the system AMD. Many of these configurations are
likely unviable due to a destabilization of the system, but a subset
may provide the seed for a scattering event that resulted in the
present system architecture \citep{kane2023b}.

The concentration of the system AMD in the outer planet is consistent
with a history in which the present architecture was dynamically
excited after formation, whether by scattering or by another secular
process, although the AMD alone does not uniquely identify the
excitation origin \citep{laskar2017}. Meanwhile, the wide
spacing of the planets implies that the system remains AMD-stable in
the sense of \citet{laskar2017} and \citet{petit2017a}. Despite the
substantial AMD stored in planet~d, there is insufficient angular
momentum exchange available to drive orbit crossing between adjacent
planets. This is consistent with the long-term N-body integrations
described in Section~\ref{sec:dyn}, which show stability for all
tested inclinations.


\subsection{Secular Excitation from an Outer Companion}
\label{sec:outer}

An unseen outer companion on a moderately eccentric orbit could excite
the eccentricity of planet~d through secular perturbations
\citep{kane2014b,petrovich2015b,anderson2017d}. \citet{nari2025}
reported sensitivity to companions with true masses of
$\sim$50~$M_\oplus$ at orbital distances of 3--10~AU, and found no
evidence for additional Keplerian signals with periods exceeding
2000~d or a long-term acceleration trend. This effectively rules out
Neptune-mass or larger companions within $\sim$10~AU on favorably
inclined orbits. However, a giant planet at larger separations
(10--20~AU) could still be present below the current detection
threshold, particularly if it resides on a low-inclination orbit that
minimizes the RV signal.
 
An additional configuration that is particularly difficult to exclude
is one in which the outer companion itself resides on a highly
eccentric orbit. The detectability of a Keplerian signal in RV data
depends sensitively on the orbital eccentricity, with the induced
stellar reflex motion becoming increasingly concentrated near the
brief periastron passage as $e$ increases
\citep{cumming2004,kane2007b,kane2013c}. For sufficiently high
eccentricities, the bulk of the orbit is spent near apastron where the
RV amplitude is substantially reduced, such that a companion may evade
detection unless the observational baseline captures a periastron
passage. This mechanism is particularly attractive in the context of
HD~20794~d for two reasons. First, it naturally reconciles the absence
of a long-term RV trend with the existence of a sufficiently massive
outer perturber, as the time-averaged reflex motion of the host star
would be small compared to the peak amplitude near periastron. Second,
a highly eccentric outer companion is a natural source of AMD, which
would be exchanged with planet~d through secular coupling and could
account for the observed eccentricity without requiring a scattering
event in the system's dynamical history
\citep{nagasawa2008,naoz2013b}. A companion with $e \gtrsim 0.5$
at semi-major axes of 10--30~AU would be consistent with the current
RV non-detection while also providing the secular forcing necessary to
maintain the eccentricity of planet~d over gigayear timescales.
 
The Herschel-detected debris disk at $\sim$24~AU \citep{kennedy2015d}
may provide indirect evidence for such a companion, as debris disk
structure is often shaped by the gravitational influence of nearby
planets \citep{wyatt2012}. Future observations with longer temporal
baselines and astrometric constraints from Gaia will be essential for
testing this hypothesis \citep{perryman2014c,holl2023a}.


\section{Discussion}
\label{sec:disc}


\subsection{Habitability Implications}
\label{sec:habitability}

The dynamical analysis presented here reveals a complex picture for
the habitability of the HD~20794 system. The system is dynamically
stable at all tested inclinations, meaning that planet~d retains its
orbit regardless of its true mass, and the inner planets are not
disrupted even in the most extreme mass scenarios. This stability
ensures that the HZ of the system is dynamically accessible, a key
prerequisite for habitability \citep{kane2024d,kane2024e}.

If planet~d is indeed a $\sim$6~$M_\oplus$ rocky world (i.e., the
system is viewed near edge-on), then its eccentric orbit carries it
through the HZ with a time-averaged flux consistent with temperate
conditions. However, the large eccentricity produces extreme seasonal
variations in incident flux, with a factor of $\sim$7 difference
between periastron and apastron. Climate modeling of Earth-like
planets on eccentric orbits suggests that atmospheres with sufficient
thermal inertia (e.g., those with substantial CO$_2$ inventories or
deep oceans) can buffer the extreme flux variations and maintain
above-freezing temperatures for a significant fraction of the orbit,
even for eccentricities as large as 0.4--0.5
\citep{williams2002,dressing2010,way2017a}. The orbit-averaged flux
of $\langle F \rangle \approx 0.42$~$F_\oplus$ for HD~20794~d is well
within the range explored by such studies, suggesting that habitable
conditions are at least plausible.

Three-dimensional general circulation model (GCM) simulations provide
the most detailed predictions for the climate response to eccentric
orbits
\citep{shields2016a,way2017a,wolf2017a,way2023a,leconte2013c}. For an
Earth analog with $e = 0.4$, \citet{way2017a} found that the ocean
thermal inertia is sufficient to prevent complete surface freezing
during apastron passage, though significant ice-sheet advance and
retreat occurs on orbital timescales. The 647.6~d orbital period of
HD~20794~d is long enough that such ``flash-freezing'' cycles would be
driven primarily by the atmospheric response time rather than the
ocean's deeper thermal reservoir. The relatively low orbit-averaged
flux ($\sim$0.42~$F_\oplus$) places HD~20794~d near the outer edge of
the conservative HZ, where the climate sensitivity to CO$_2$ partial
pressure becomes a critical factor
\citep{kopparapu2013a,shields2016a}. A dedicated GCM study of
HD~20794~d, incorporating the measured orbital parameters and a range
of atmospheric compositions, would be a valuable follow-up to the
dynamical analysis presented here.

If the true mass of planet~d is significantly higher than the minimum
mass, for example $\sim$10--20~$M_\oplus$ corresponding to
inclinations of $\sim$17--34\degr, then the planet would more likely
be a mini-Neptune with a thick volatile envelope
\citep{rogers2015a,otegi2020,muller2024b}, rendering it inhospitable
to Earth-based life. Moreover, at such masses the eccentric orbit of
planet~d would exert a strong dynamical influence on any terrestrial
planet residing elsewhere within the HZ. Previous dynamical studies
have shown that even a mildly eccentric sub-Neptune in or near the HZ
can destabilize the orbits of additional temperate planets in the same
system, substantially reducing the phase space within which habitable
worlds may reside \citep{kane2023d,kane2024e}. In the case of
HD~20794, the periastron of planet~d (0.74~AU) lies deep within the
HZ, such that gravitational perturbations from planet~d would be
strongly felt by any additional HZ bodies. The combination of an
eccentric orbit, a potentially substantial mass, and a periastron
passage within the HZ therefore makes planet~d a potential dynamical
disruptor for any other HZ planet in the system, regardless of whether
planet~d itself is rocky or volatile-rich. This reinforces the
importance of the current three-planet architecture: if planet~d is
habitable, it is likely the sole habitable world in the HD~20794
system, with the notable exception of subsurface ocean
  environments analogous to those hypothesized for the Jovian and
  Saturnian satellites \citep{lunine2017}.


\subsection{Relevance to HWO and Future Direct Imaging}
\label{sec:imaging}

\begin{figure*}
  \begin{center}
    \begin{tabular}{cc}
      \includegraphics[angle=270,width=\columnwidth]{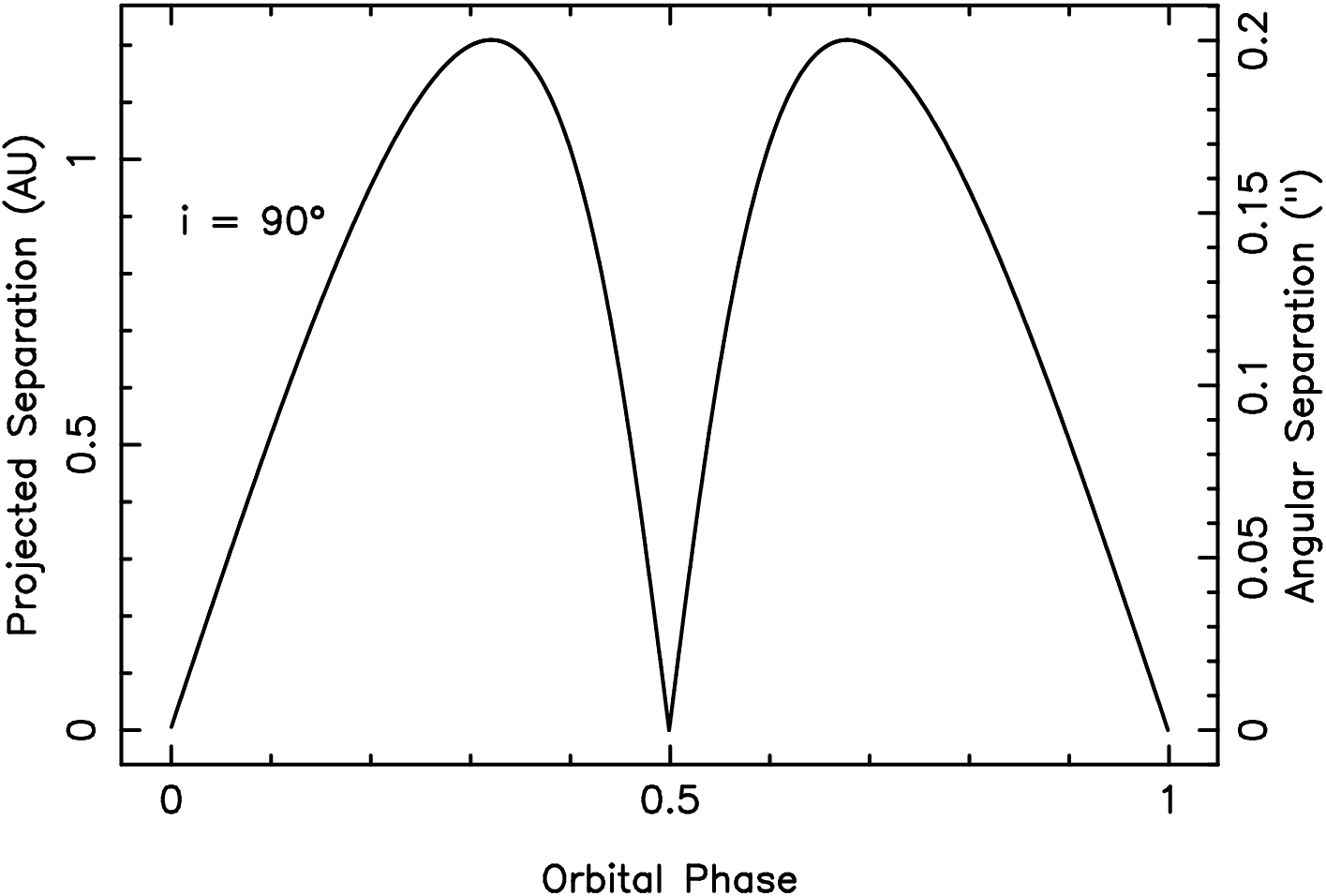} &
      \includegraphics[angle=270,width=\columnwidth]{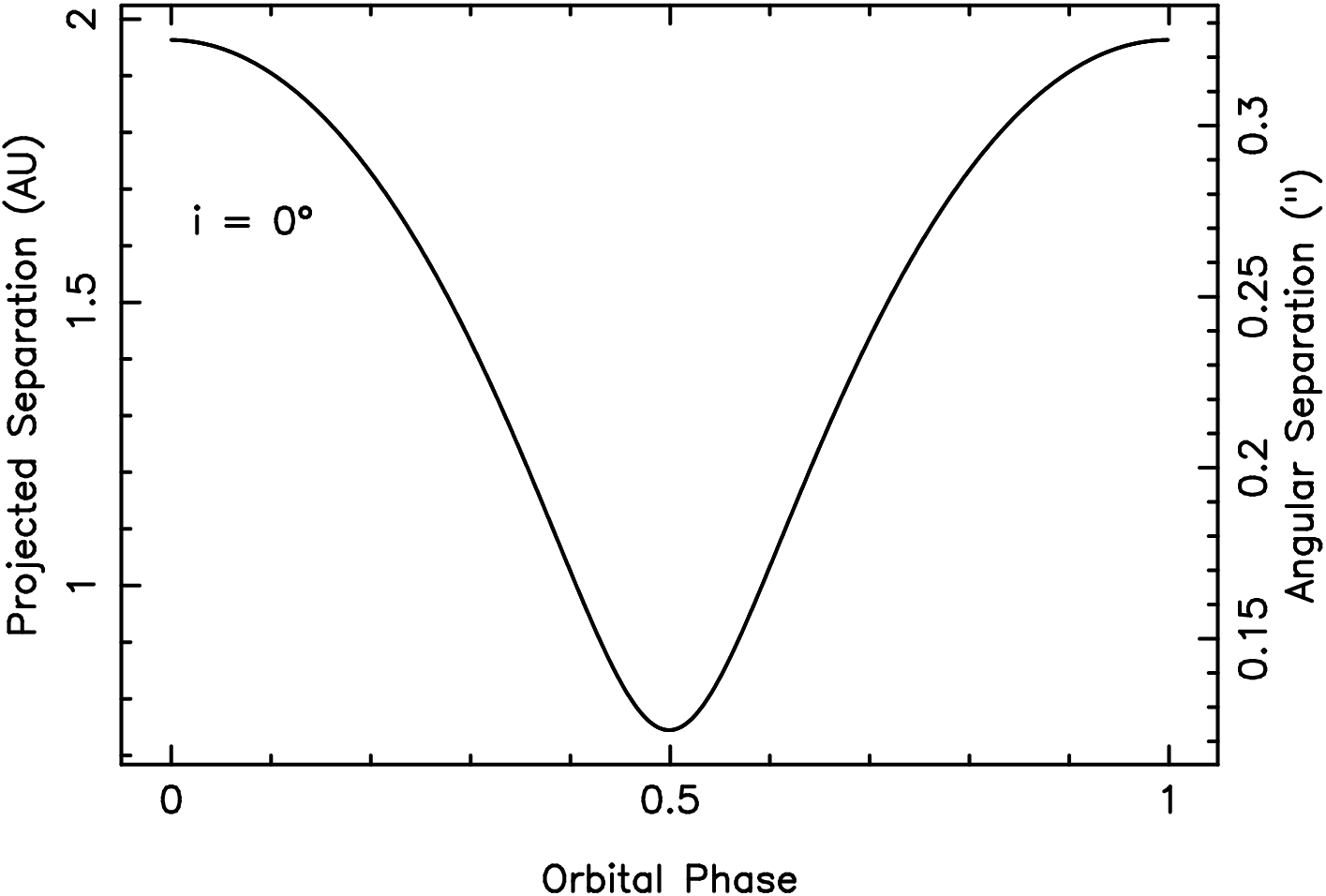}
    \end{tabular}
  \end{center}
  \caption{Projected and angular separation of HD~20794~d from the
    host star, assuming inclinations of $i = 90\degr$ (left) and $i =
    0\degr$ (right). An orbital phase of zero corresponds to the
    location of superior conjunction.}
  \label{fig:sep}
\end{figure*}

HD~20794 is included on the HWO ExEP precursor science stars list
\citep{mamajek2024}, and the proximity of the system ($d = 6.04$~pc)
makes it exceptionally favorable for direct imaging characterization
at visible and near-infrared wavelengths \citep{nari2025}. Shown in
Figure~\ref{fig:sep} are the predicted projected and angular
separations of HD~20794~d from the host star over one complete
orbit. The two panels of Figure~\ref{fig:sep} represent the extreme
inclination scenarios of $i = 90\degr$ (left) and $i = 0\degr$
(right), the latter of which corresponds to a face-on orbit. The
angular separation of planet~d from the host star depends on the
inclination, ranging from $\sim$0.0--0.20\arcsec for the $i = 90\degr$
case and $\sim$0.12--0.33\arcsec for the $i = 0\degr$ case, placing it
within the accessible regime of a 6-meter class space telescope with a
coronagraph \citep{harada2024b}. Additionally, \citet{laliotis2023}
provided RV sensitivity constraints for HD~20794 in the context of
future direct imaging missions, further underscoring the importance of
a comprehensive dynamical characterization. Systems in which the known
planets are stable and the HZ is dynamically accessible are preferred
targets for atmospheric characterization \citep{kane2024d,kane2024e}.
The dynamical stability demonstrated here at all tested inclinations
strengthens the case for HD~20794 as a high-priority target, with
planet~d itself representing the most promising HZ candidate in the
system given the likely dynamical constraints on additional HZ bodies
(Section~\ref{sec:habitability}).

The HD~20794 system is also a compelling target for the Large
Interferometer for Exoplanets (LIFE) mission concept
\citep{quanz2022a}, which would characterize terrestrial exoplanet
atmospheres in the mid-infrared through space-based nulling
interferometry. The mid-infrared offers
complementary atmospheric diagnostics to those accessible at visible
wavelengths, including thermal emission features of CO$_2$, O$_3$, and
H$_2$O that are critical biosignature and habitability indicators
\citep{schwieterman2018,quanz2022b,fujii2018}. The combination of HWO
reflected-light and LIFE thermal-emission observations would provide a
comprehensive atmospheric characterization of planet~d
\citep{fujii2018,catling2018,alei2024}, enabling robust constraints on
its surface temperature, atmospheric composition, and potential
habitability across the full range of its eccentric orbit.


\subsection{Comparison to Other Eccentric HZ Planets}
\label{sec:comparison}

The majority of known planets on eccentric orbits that traverse the HZ
are giant planets with masses of order $\sim$$M_J$. These include
HR~5183~b \citep{blunt2019,kane2019b}, the recently characterized
companions in the Gaia-4 and Gaia-5 systems
\citep{stefansson2025,kane2025c}, and numerous other cases cataloged
through RV surveys \citep{kane2012d,kane2012e,kane2021a}. For such
massive planets, habitability of the planet itself is not plausible,
but the orbital dynamics have implications for any terrestrial bodies
that may share the HZ. HD~20794~d occupies a fundamentally different
regime. At $M_p \sin i = 5.82$~$M_\oplus$ with $e = 0.45$, it is the
lowest (minimum) mass confirmed planet with an eccentricity exceeding
0.4 whose orbit crosses the HZ of its host star. This combination of
low mass and high eccentricity is exceedingly rare among the known
exoplanet population: statistical studies of the eccentricity
distribution consistently find that super-Earths and sub-Neptunes
($M_p \lesssim 20$~$M_\oplus$) reside on near-circular orbits, with
mean eccentricities of $\langle e \rangle \approx 0.04$--0.06 for
compact multi-planet systems \citep{shen2008c,kane2012d}. Fewer than
five confirmed planets below 20~$M_\oplus$ have eccentricities
exceeding 0.4.

The closest comparator in the literature is Gl~514~b, a super-Earth
with $M_p \sin i \approx 5.2$~$M_\oplus$ on a similarly eccentric
orbit ($e \approx 0.45$) that crosses the HZ of its M-dwarf host at
7.6~pc \citep{damasso2022}. \citet{biasiotti2024} conducted a
comprehensive climate study of Gl~514~b using the EOS-ESTM
seasonal-latitudinal energy balance model, demonstrating that
habitability remains achievable for plausible atmospheric compositions
despite the large flux variations, but is sensitive to CO$_2$
abundance, rotation period, and ocean fraction. A comparable climate
investigation of HD~20794~d, incorporating the measured orbital
parameters and a range of atmospheric and surface configurations,
would represent a natural extension of the Gl~514~b study to the
G-dwarf regime. Despite the similarities in mass and eccentricity
between Gl~514~b and HD~20794~d, the two planets occupy quite
different stellar environments: Gl~514~b orbits an M dwarf, where the
XUV environment, tidal effects, and spectral energy distribution
differ substantially from the G-dwarf host of HD~20794. HD~20794~d is
therefore the only confirmed super-Earth on a highly eccentric
HZ-crossing orbit around a Sun-like star within 10~pc.

The rarity of this parameter combination underscores its scientific
value. The high eccentricity of planet~d cannot be easily explained by
the dynamical interactions among the three known planets alone
(Section~\ref{sec:scattering}), and the low mass makes conventional
planet-planet scattering an unlikely formation pathway. The proximity
of the host star ($d = 6.04$~pc), its inclusion on the HWO precursor
target list \citep{mamajek2024,tuchow2024,harada2024b}, and the
availability of a Sun-like stellar spectrum for atmospheric retrieval
modeling make HD~20794~d an important test case for understanding
whether habitable conditions can persist on planets with large orbital
eccentricities around solar-type stars.


\section{Conclusions}
\label{sec:con}

The HD~20794 system provides a compelling laboratory for studying the
relationship between orbital dynamics and planetary habitability. The
CHZ spans 0.807--1.440~AU, and HD~20794~d spends approximately 32\% of
its orbital period within this zone, with an orbit-averaged flux of
$\langle F \rangle \approx 0.42$~$F_\oplus$ consistent with temperate
conditions for an Earth-like atmosphere.  N-body simulations reveal
that the three-planet system is dynamically stable at all tested
inclinations, from $i = 90\degr$ ($M_d \approx 5.8$~$M_\oplus$) down
to $i = 5\degr$ ($M_d \approx 66.8$~$M_\oplus$), over the full
$10^7$~year integration. The eccentricities of all three planets
undergo coupled secular oscillations whose eigenperiod scales
inversely with the total system mass, in precise agreement with
Laplace-Lagrange secular theory. The eccentricity of planet~d remains
tightly confined near its initial value of $e \approx 0.45$, while the
inner planets exhibit forced eccentricity amplitudes ($\Delta e_b
\approx 0.05$, $\Delta e_c \approx 0.02$) that are consistent with the
measured eccentricities at all inclinations. The true planetary masses
are therefore not constrained by the dynamical stability of the
system, but may be probed by future high-precision eccentricity
measurements of the inner planets. The origin of the eccentricity of
planet~d remains uncertain, with planet-planet scattering, secular
excitation from an unseen outer companion, and primordial eccentricity
all viable mechanisms that require further observational constraints.

The proximity of HD~20794 ($d = 6.04$~pc) and its inclusion on the HWO
precursor target list make this system a high-priority case for
understanding the dynamical prerequisites for habitability in
multi-planet systems. The dynamical stability demonstrated here at all
inclinations strengthens the case for HD~20794 as a viable direct
imaging target, and the system is also compelling for the LIFE
mid-infrared interferometry concept. We note, however, that the
eccentric orbit of planet~d carries its periastron deep into the HZ,
such that planet~d is likely to act as a dynamical disruptor for any
additional terrestrial planets within the HZ
\citep{kane2023d,kane2024e}. Planet~d therefore represents the
dominant HZ target in the system, and its habitability is not
contingent on a companion population of temperate worlds. Future work
should incorporate three-dimensional GCM simulations to quantify the
surface temperature evolution of HD~20794~d over its eccentric orbit,
and further investigation of the debris disk at $\sim$24~AU may
illuminate the dynamical history of the outer system and the prospects
for volatile delivery to the inner planets.


\section*{Acknowledgements}

The author would like to thank the anonymous reviewer for their
  useful feedback on the manuscript. This research has made use of
the Habitable Zone Gallery at hzgallery.org. The results reported
herein benefited from collaborations and/or information exchange
within NASA's Nexus for Exoplanet System Science (NExSS) research
coordination network sponsored by NASA's Science Mission Directorate.


\software{Mercury \citep{chambers1999}}




\end{document}